\documentclass[aps,prd,twocolumn,floatfix,nofootinbib,superscriptaddress]{revtex4-1}
\usepackage[colorlinks,linkcolor=blue,anchorcolor=blue,citecolor=blue,urlcolor=blue,breaklinks=true]{hyperref}
\usepackage[libertine]{newtxmath}
\usepackage{graphicx}
\usepackage{slashed}
\usepackage{amsfonts}
\usepackage{amsmath}
\usepackage{blindtext}
\usepackage{microtype}
\usepackage{easyReview}
\usepackage{booktabs}
\usepackage{cleveref}
\usepackage{eucal}

\newcommand{\vect}[1]{\boldsymbol{#1}}

\graphicspath{{.}{figures/}}

\begin{document}

\title{Unpolarized generalized parton distributions of light and heavy vector mesons.}

\author{Chao Shi}
\email[]{cshi@nuaa.edu.cn}
\affiliation{Department of Nuclear Science and Technology, Nanjing University of Aeronautics and Astronautics, Nanjing 210016, China}

\author{Jicheng Li}
\affiliation{Department of Nuclear Science and Technology, Nanjing University of Aeronautics and Astronautics, Nanjing 210016, China}

\author{Pei-Lin Yin}
\affiliation{School of Science, Nanjing University of Posts and Telecommunications, Nanjing 210023, China}

\author{Wenbao Jia}
\affiliation{Department of Nuclear Science and Technology, Nanjing University of Aeronautics and Astronautics, Nanjing 210016, China}

\begin{abstract}
We study the leading-twist unpolarized generalized parton distributions (GPDs) of light and heavy vector mesons, i.e., the $\rho$, $J/\psi$ and $\Upsilon$, at zero skewness. An ansatz incorporating the zero mode contribution is introduced to modify the light front overlap representation of GPDs. The leading Fock-state light front wave functions (LF-LFWFs) of vector mesons from DS-BSEs approach are then employed to study the meson GPDs. The light front spatial distribution of valence quarks within vector mesons is then studied with the impact parameter dependent GPD (IPD GPD). We also investigate the electromagnetic and gravitational form factors, which are the first and second Mellin moments of the GPDs. The light-cone mass radius of $\rho$ is determined to be $0.30$ fm, close to a recent NJL model prediction $0.32$fm. For $J/\psi$ and $\Upsilon$, they are predicted to be $0.151$fm and $0.089$ fm respectively.  
\end{abstract}
\maketitle

%===============================================================================
%===============================================================================
\section{INTRODUCTION\label{intro}}
The generalized parton distribution functions (GPDs) extend the 1-dimensional collinear parton distribution functions (PDFs) to the 3-dimensional case, thus allowing a femtoscale tomography of the hadrons and nuclei \cite{Ji:1996nm,Radyushkin:1997ki,Burkardt:2000za,Burkardt:2002hr}. Meanwhile, it builds a direct connection between hadrons' electromagnetic and mechanical properties with their partonic substructure \cite{Diehl:2003ny,Belitsky:2005qn,Polyakov:2002yz,Polyakov:2018zvc,Lorce:2018egm}. Experimentally, the GPDs are accessible in various hard exclusive processes, such as deeply virtual Compton scattering (DVCS), deeply virtual meson production (DVMP) and time-like Compton scattering (TCS) \cite{Collins:1996fb,Ji:1996nm,Radyushkin:1997ki,Berger:2001xd}. Interest from both theory and experimental sides thus motivates next generation facilities as the electron-ion colliders \cite{AbdulKhalek:2021gbh,Anderle:2021wcy,LHeC:2020van}.

Despite great interest resides in spin-0 and spin-1/2 targets, the general formalism of unpolarized and polarized GPDs of spin-1 target was investigated for the case of deuteron \cite{Berger:2001zb,Cosyn:2018rdm}, followed by various model calculations \cite{Cano:2003ju,Dong:2013rk,Mondal:2017lph,Cosyn:2018rdm}. Meanwhile, the GPDs of vector mesons were studied by various light front quark models \cite{Sun:2017gtz,Sun:2018ldr,Kumar:2019eck,Adhikari:2018umb} and the Nambu--Jona-Lasinio model (NJL) model \cite{Zhang:2022zim}. In recent years, with the growing interest in gravitational form factors, the GPDs of vector mesons, whose second Mellin moments yield the GFFs, provide an important handle to study the mechanical properties of vector mesons \cite{Polyakov:2019lbq,Cosyn:2019aio}. The $\rho$ GPDs are also connected with the generalized distribution amplitudes (GDAs), with the latter being the analytic continuation of GPDs to the crossed channel \cite{Teryaev:2001qm,Diehl:2003ny,Kumano:2017lhr}. Experimentally the information of GDAs are accessible in exclusive process $\gamma \gamma^*\rightarrow \rho \rho$ \cite{Anikin:2003fr,Anikin:2005ur}.

In this work, we present a model calculation of the unpolarized GPDs of vector mesons through a synergy between the DS-BSEs and the light front approaches. The DS-BSEs approach has a long history of successfully predicting various meson and baryon properties \cite{Roberts:2007ji,Cloet:2013jya,Frederico:2013vga,Eichmann:2016yit,Yin:2019bxe,dePaula:2020qna}. Regarding the vector mesons, the $\rho$ mass and decay constant were first predicted with the Maris-Tandy model in \cite{Maris:1999nt}, and then a fully covariant calculation on EMFFs of $\rho$ and $J/\psi$ in the instant space-time form was given in \cite{Bhagwat:2006pu,Maris:2006ea}. It is certainly desirable to generalize such fully covariant calculation to the case of GPDs. However, technical difficulties exist at present. Here we resort to the light front overlap representation, and use the BSEs-based LF-LFWFs as input \cite{Shi:2021taf,Shi:2022erw}. In this regard, the nonperturbative dynamical information of vector mesons is conveyed from DS-BSEs to the GPDs. So this work presents an initial effort from DS-BSEs toward the  vector meson GPDs. Moreover, as the light and heavy vector mesons can be simultaneously studied with the same truncation scheme in the DS-BSEs formalism, it also provides a good opportunity to see how the GPDs evolve as the current mass of the valence quark  increases. Physically, this is accompanied by the diminishing of dynamical chiral symmetry breaking, as well as the relativistic effect.

This paper is organized as follows. In section \ref{sec:overlap} we recapitulate the general formalism of vector meson GPDs and their overlap representation, and also the BSEs-based LF-LFWFs of vector mesons. To incorporate the zero mode contribution, a revised ansatz of GPD overlap representation is proposed. In section \ref{sec:den}, we first show 3-dimensional distribution with the help of IPD GPDs. The electromagnetic form factors and multipole moments, as well as certain gravitational form factors and  light-cone mass radius are then given. We finally summarize in section \ref{sec:sum}.

\section{Unpolarized GPDs of vector meson \label{sec:overlap}}

In the light-cone gauge, the unpolarized quark GPDs of spin-1 hadrons are defined through the correlation function 
\begin{align}
V_{\Lambda',\Lambda}(x,\xi,t) =\int \frac{dz^-}{4 \pi} e^{i x P^+ z^-}\left \langle p', \Lambda' \left | \bar{\psi}\left(-\frac{z^-}{2}\right)\gamma^+ \psi \left(\frac{z^-}{2}\right)\right | p,\Lambda \right \rangle 
\label{eq:Vdef}
\end{align}
Here the $p$ and $p'$ are the four-momentum of incoming and outgoing hadrons, with $\Lambda$ and $\Lambda'=0,\pm 1$ denoting their helicity. The light front vector definition takes the convention $a^{\pm}=(a^0 \pm a^3)/\sqrt{2}$ and the light front four vector thus is $a^\mu=(a^+,a^-,\vect{a}_\perp)$. Other variables used are $P^\mu=(p'^\mu+p^\mu)/2$, $\Delta^\mu=p'^\mu-p^\mu$, $t=\Delta^2$ and skewness variable $\xi=-\Delta^+/(2 P^+)$. At leading twist, there are five GPDs that enter the decomposition of $V_{\Lambda',\Lambda}$ \cite{Berger:2001zb}
\begin{widetext}
\begin{align}
V_{\Lambda',\Lambda}(x,\xi,t)&=-(\epsilon'^*\cdot \epsilon) H_1+\frac{(\epsilon \cdot n)(\epsilon' \cdot P)+(\epsilon' \cdot n)(\epsilon \cdot P)}{P \cdot n}H_2-2\frac{(\epsilon \cdot P)(\epsilon'^* \cdot P)}{M^2}H_3+\frac{(\epsilon \cdot n)(\epsilon' \cdot P)-(\epsilon'^* \cdot n)(\epsilon \cdot P)}{P\cdot n}H_4 \nonumber\\ 
&+\left\{M^2\frac{(\epsilon \cdot n)(\epsilon'^* \cdot n)}{(P\cdot n)^2}+\frac{1}{3}(\epsilon'^* \cdot \epsilon)\right\} H_5. \label{eq:Vdeco}
\end{align}
\end{widetext}
The polarization vector $\epsilon \equiv \epsilon^\mu(p,\Lambda)$ and  $\epsilon' \equiv \epsilon^\mu(p',\Lambda')$. Parity and time reversal invariance then lead to \cite{Berger:2001zb,Cano:2003ju}
\begin{align}
V_{\Lambda',\Lambda}(x,\xi,t)&=(-1)^{\Lambda'-\Lambda} V_{-\Lambda',-\Lambda}(x,\xi,t), \label{eq:PI}\\
V_{\Lambda',\Lambda}(x,\xi,t)&=(-1)^{\Lambda'-\Lambda} V_{\Lambda,\Lambda'}(x,-\xi,t). \label{eq:TI}
\end{align}
Using Eq.~(\ref{eq:PI}), five $V_{\Lambda',\Lambda}$'s are independent, e.g.,  $V_{0,0}$, $V_{0,1}$, $V_{1,0}$, $V_{1,1}$ and $V_{1,-1}$. Further, at zero skewness, $V_{0,1}$ and $V_{1,0}$ are related with Eq.~(\ref{eq:TI}). In the end, for our purpose of studying $H_i$ at zero skewness, there are four independent $V_{\Lambda',\Lambda}$'s left, which will be taken as $V_{0,0}$, $V_{0,1}$, $V_{1,1}$ and $V_{1,-1}$ in the following. To reverse Eq.~(\ref{eq:Vdeco}), one can take a specific Breit frame ($\vect{p}+\vect{p'}=0$) in which the four-vectors are  \cite{Bakker:2002mt,Adhikari:2018umb}
\begin{align}
n^{\mu}&=(0,\sqrt{2},0,0)\\
\Delta^\mu&=(0,0,|\vect{\Delta}_\perp|,0),\\
p^\mu&=\left(M\frac{\sqrt{1+\tau}}{\sqrt{2}}, M\frac{\sqrt{1+\tau}}{\sqrt{2}}, -\frac{|\vect{\Delta}_\perp|}{2},0 \right),\\
p'^\mu&=\left(M\frac{\sqrt{1+\tau}}{\sqrt{2}}, M\frac{\sqrt{1+\tau}}{\sqrt{2}}, \frac{|\vect{\Delta}_\perp|}{2},0 \right),
\end{align}

with $\tau=\vect{\Delta}^2_\perp/(4 M^2)$, and the polarization vector reads
\begin{align}
\epsilon^\mu(p,\Lambda=\pm 1)&=\mp \frac{1}{\sqrt{2}}\left(0,-\frac{|\vect{\Delta}_\perp|}{\sqrt{2}p^+},1,\pm i \right), \\
\epsilon^\mu(p',\Lambda'=\pm 1)&=\mp \frac{1}{\sqrt{2}}\left(0,\frac{|\vect{\Delta}_\perp|}{\sqrt{2}p^+},1,\pm i \right), \\
\epsilon^\mu(p,\Lambda=0)&=\frac{1}{M}\left(\frac{p^+}{\sqrt{2}},\frac{-M^2+\vect{\Delta}_\perp^2/4}{\sqrt{2} p^+},-\frac{|\vect{\Delta}_\perp|}{2},0 \right), \\
\epsilon^\mu(p',\Lambda'=0)&=\frac{1}{M}\left(\frac{p^+}{\sqrt{2}},\frac{-M^2+\vect{\Delta}_\perp^2/4}{\sqrt{2} p^+},\frac{|\vect{\Delta}_\perp|}{2},0 \right). 
\end{align}

Reversing Eq.~(\ref{eq:Vdeco})  yields 
\begin{align}
H_1&= \frac{1}{3}[V_{0,0}-2(\tau-1)V_{1,1} 
+2\sqrt{2\tau}V_{1,0}+2V_{1,-1}],\label{eq:H1}\\
H_2 &= 2V_{1,1} -\frac{2}{\sqrt{2\tau}} V_{1,0},\label{eq:H2}\\
H_3&=-\frac{V_{1,-1}}{\tau},\label{eq:H3}\\
H_4&=0,\\
H_5&= V_{0,0}-(1+2\tau)V_{1,1}+2\sqrt{2\tau}V_{1,0}-V_{1,-1}, \label{eq:H5}
\end{align}
with the abbreviation $H_i=H_i(x,0,t)$ and $V_{\Lambda',\Lambda}=V_{\Lambda',\Lambda}(x,0,t)$.

The light front overlap representation of correlation function $V_{\Lambda',\Lambda}$ can be obtained using the Fock state expansion of meson state  and  canonical expansion of the (anti)quark field. Its final form reads \cite{Diehl:2003ny,Adhikari:2018umb},
\begin{align}
V_{\Lambda',\Lambda}(x,0,t;\mu_0)=\sum_{\lambda_q,\lambda_{\bar{q}}}\int \frac{d^2\vect{k}_T}{2 (2\pi)^3}  \Phi^{\Lambda' *}_{\lambda_q,\lambda_{\bar{q}}}(x,\hat{\vect{k}}_T)  \Phi^{\Lambda}_{\lambda_q,\lambda_{\bar{q}}}(x,\tilde{\vect{k}}_T),
\label{eq:GPD}
\end{align}
with $\hat{\vect{k}}_T=\vect{k}_T+(1-x)\frac{\vect{\Delta}_T}{2}$ and $\tilde{\vect{k}}_T=\vect{k}_T-(1-x)\frac{\vect{\Delta}_T}{2}$. The  $\Phi^{\Lambda}_{\lambda_q,\lambda_{\bar{q}}}(x,\vect{k}_T)$ is the LFWF of $q\bar{q}$-component of the vector meson. 

To proceed, we take the $\rho^+$, $J/\psi$ and $\Upsilon$ LF-LFWFs obtained with the DS-BSEs approach \cite{Shi:2021taf,Shi:2022erw}. To be more specific, we first numerically solve the quark propagator $S(p)$ and vector meson Bethe-Salpeter amplitudes $\Gamma_\mu(k,P)$ in the Rainbow-Ladder truncation. Then, using the projection formula
\begin{align}\label{eq:chi2phi}
	\Phi^\Lambda_{\Lambda,\Lambda'}(x,\vect{k}_T)&=-\frac{1}{2\sqrt{3}}\int \frac{dk^- dk^+}{2 \pi} \delta(x P^+-k^+) \nonumber\\
	&\hspace{20mm}\textrm{Tr}\left [\Gamma_{\Lambda,\Lambda'}\gamma^+ \chi^M(k,P) \cdot \epsilon_\Lambda(P) \right ],
\end{align}
the covariant BS wave functions are projected onto the light front and the LF-LFWFs are obtained. Here the $\chi^M_\mu(k,P)=S(k+\eta P)\Gamma^M_\mu(k,P)S(k-(1-\eta)P) $ and the $\epsilon_\Lambda(P)$ is the  meson polarization vector. The $\Gamma_{\pm,\mp}=I\pm \gamma_5$ and $\Gamma_{\pm,\pm}=\mp(\gamma^1\mp i\gamma^2)$  correspond to different quark-antiquark helicity configurations. The trace is taken over Dirac, color and flavor spaces.  The obtained LF-LFWFs well produced diffractive electroproduction vector meson data at HERA within the color dipole picture \cite{Shi:2021taf}, and yield novel results on vector meson TMDs that are sensitive to higher orbital angular momentum \cite{Shi:2022erw}.

Since Eq.~(\ref{eq:GPD}) has taken the leading Fock-state truncation,  the BSEs-based LF-LFWFs are further rescaled to satisfy the normalization condition
\begin{align}
	1&=\sum_{\lambda_q,\lambda_{\bar{q}}} \int_0^1 dx \int \frac{d \vect{k}_T^2}{2(2 \pi)^3}  |\Phi^{\Lambda,(\textrm{re})}_{\lambda_q,\lambda_{\bar{q}}}(x,\vect{k_T})|^2. \label{eq:N2},
\end{align}
with $\Phi_{\Lambda,\Lambda'}^{\Lambda=0,(\rm{re})}\! =N_1 \Phi_{\Lambda,\Lambda'}^{\Lambda=0}$ and $\Phi_{\Lambda,\Lambda'}^{\Lambda=\pm 1,(\rm{re})}\! =N_2 \Phi_{\Lambda,\Lambda'}^{\Lambda=\pm 1}$. This ensures the quark number sum rule for vector mesons with $\Lambda=0,\pm 1$ respectively 
\begin{align}
\int_0^1 dx f_\Lambda(x)\equiv \int_0^1 dx V_{\Lambda,\Lambda}(x,0,0)=1, \label{eq:norm1}
\end{align}  
where the $f_\Lambda(x)$ is the collinear distribution of unpolarized quark in vector meson with helicity $\Lambda$.

At this stage, it seems straightforward to calculate the $H_i$ ($i=1,2,3,5$) of vector mesons using Eqs.~(\ref{eq:H1}-\ref{eq:GPD}). However, as we have noticed, there are two sources of zero-mode contributions encountered in the literatures regarding the vector meson GPDs. The first one is found in the calculation of vector meson EMFF with the  triangle diagram using bare photon-quark vertex $\gamma^\mu$ \cite{Bakker:2002mt,Choi:2004ww}. The authors demonstrated with explicit calculation that $I_{0,0}(t)=\int dx V_{0,0}(x,0,t)$ in the front form receives nonvalence contribution, which is referred to as the zero-mode contribution since it originates in the nonvalence region which shrinks to zero in the limit  $p^+ \rightarrow p'^+$, or namely $\Delta^+ \rightarrow 0$. In terms of GPD, this suggests that there is a nontrivial Efremov-Radyushkin-Brodsky-Lepage (ERBL) region \cite{Efremov:1979qk,Lepage:1979zb} contribution in $V_{0,0}(x,0,t)$, which doesn't vanish in the limit $\xi \rightarrow 0$, but rather yields a finite contribution when integrated over $x$. Analytically, this property can be realized with an ansatz as $V'_{0,0}(x,0,t)=V_{0,0}(x,0,t)+\bar{F}(t)\delta(x)$, where the second term mimics the zero-mode modification. It should be emphasized that $V_{\Lambda,\Lambda'}$'s with helicity configurations $\Lambda\ne 0$ or $\Lambda'\ne 0$ are free of such zero-mode contribution \cite{Bakker:2002mt}.

On the other hand, an NJL model calculation of GPD using the triangle diagram shows zero-mode contribution could also arise when the bare photon-quark vertex gets fully dressed, i.e., from $\gamma^\mu$ to $\Gamma^\mu$. The $\Gamma^\mu$ can be obtained by solving the inhomogeneous Bethe-Salpeter equation, with $\gamma^\mu$ its inhomogeneous bare driving term. In the appendix of \cite{Shi:2020pqe}, the zero mode contribution is analytically shown to be proportional to $\delta(x)$, which serves as a \emph{hidden} ERBL region and contributes nontrivially when integrated over $x$. This zero-mode contribution originates from the dressing of photon-quark vertex, so it is independent of the type of hadron or its polarization. Therefore all the $V_{\Lambda,\Lambda'}$ should receive such contribution, which is unlike the first kind of zero-mode. Summarizing all these considerations, we eventually propose an ansatz for the modified GPDs 
\begin{align}
V'^{M}_{\Lambda,\Lambda'}(x,0,t)&=V^{M}_{\Lambda,\Lambda'}(x,0,t)+\delta_{\Lambda 0}\delta_{\Lambda' 0}\bar{F}_{M}(t)\delta(x) \nonumber \\
&+\delta(x) \tilde{F}_M(t)\int_0^1 dy V^M_{\Lambda,\Lambda'}(y,0,t). \label{eq:Vfinal} 
\end{align}
The third term on the right hand side corresponds to the second kind of zero mode contribution. The $M$ here denotes the meson dependence. As the SU(3) NJL model deals with light quarks, it only provides $\tilde{F}_{\rho}(t)$\footnote{The expression of $\tilde{F}_{\rho}(t)$ can be found in Eq.~(B13) of the appendix of  \cite{Shi:2020pqe}}, so we assume $\tilde{F}_{J/\psi}(t) =\tilde{F}_{\Upsilon}(t) \approx 0$. This is physically reasonable in the sense that the dressing effect in heavy-quark-photon vertex is more suppressed than in light-quark-photon vertex. As for the $\bar{F}_M(t)$, we will determine it using the so called angular momentum condition \cite{Grach:1983hd,Choi:2004ww}, which will be addressed in connection with the EMFFs of vector mesons later. In the following, we will use $V$ to refer to the modified $V'$ in Eq.~(\ref{eq:Vfinal}) for convenience.

\section{Density distribution and form factors \label{sec:den}}
Aligning Eqs.~(\ref{eq:H1}-\ref{eq:H5},\ref{eq:GPD},\ref{eq:Vfinal}), we can calculate all the unpolarized GPDs at zero skewness. Next we utilize these GPDs to explore various properties of the vector meson, including the 3-dimensional parton distribution and electromagnetic and gravitational form factors.
\subsection{Impact parameter dependent parton distribution function of vector mesons.}
It is well known that the GPDs encode the density distribution of quarks within hadrons in a joint space of longitudinal momentum and transverse spatial coordinate \cite{Burkardt:2000za,Burkardt:2002hr}. In the case of pion and nucleon, which are spin-0 and 1/2 respectively, it had been shown that the Fourier transform of unpolarized GPD $H(x,\xi=0,t)$ gives rise to the unpolarized GPD in the impact parameter space (IPD GPD) \cite{Burkardt:2002hr}, i.e.,
\begin{align}
\rho(x,\vect{b}_\perp)=\int \frac{d^2 \vect{\Delta}_\perp}{(2\pi)^2}H(x,0,-\Delta^2) {\textrm e}^{-i \vect{b}_\perp \cdot \vect{\Delta}_\perp}, \label{eq:rho}
\end{align}
which has the physical meaning of density distribution of unpolarized quarks within unpolarized hadron in the $x-\vect{b}_T$ space. We remind that the unpolarized impact parameter dependent PDF was originally defined as  \cite{Burkardt:2002hr}
\begin{align}
\rho_{\Lambda}(x,\vect{b}_\perp)\equiv \langle P^+,\vect{R}_\perp=\vect{0}_\perp, \Lambda| \hat{{\cal O}}_q(x,\vect{b}_\perp)| P^+,\vect{R}_\perp=\vect{0}_\perp,\Lambda \rangle,\label{eq:rhoLam}
\end{align}
with the operator 
\begin{align}
\hat{{\cal O}}_q(x,\vect{b}_\perp)&=\int \frac{dz^-}{4\pi}\bar{q}\left(-\frac{z^-}{2},\vect{b}_\perp \right)\gamma^+ q\left(\frac{z^-}{2},\vect{b}_\perp \right) \textrm{e}^{ix P^+ z^-} \\
&\sim {\cal N} \tilde{b}^\dagger (x P^+,\vect{b}_\perp)\tilde{b} (x P^+,\vect{b}_\perp)
\end{align}
characterizing the probability density of unpolarized quark at $x$ and $\vect{b}_\perp$. Here the hadron state is localized at the origin of transverse center of momentum, i.e., $\vect{R}_\perp=\sum_i x_i \vect{r}_{\perp,i}=0$. The $\Lambda$ in Eq.~(\ref{eq:rhoLam}) indicates the helicity of the hadron. For nucleon, the $\Lambda=1/2$ or $-1/2$ yield the same $\rho_{\Lambda}(x,\vect{b}_\perp)$, hence the $\Lambda-$dependence can be dropped and Eq.~(\ref{eq:rhoLam}) leads to Eq.~(\ref{eq:rho}) \cite{Burkardt:2002hr}. 

\begin{figure}[htbp]
\centering\includegraphics[width=1.0\columnwidth]{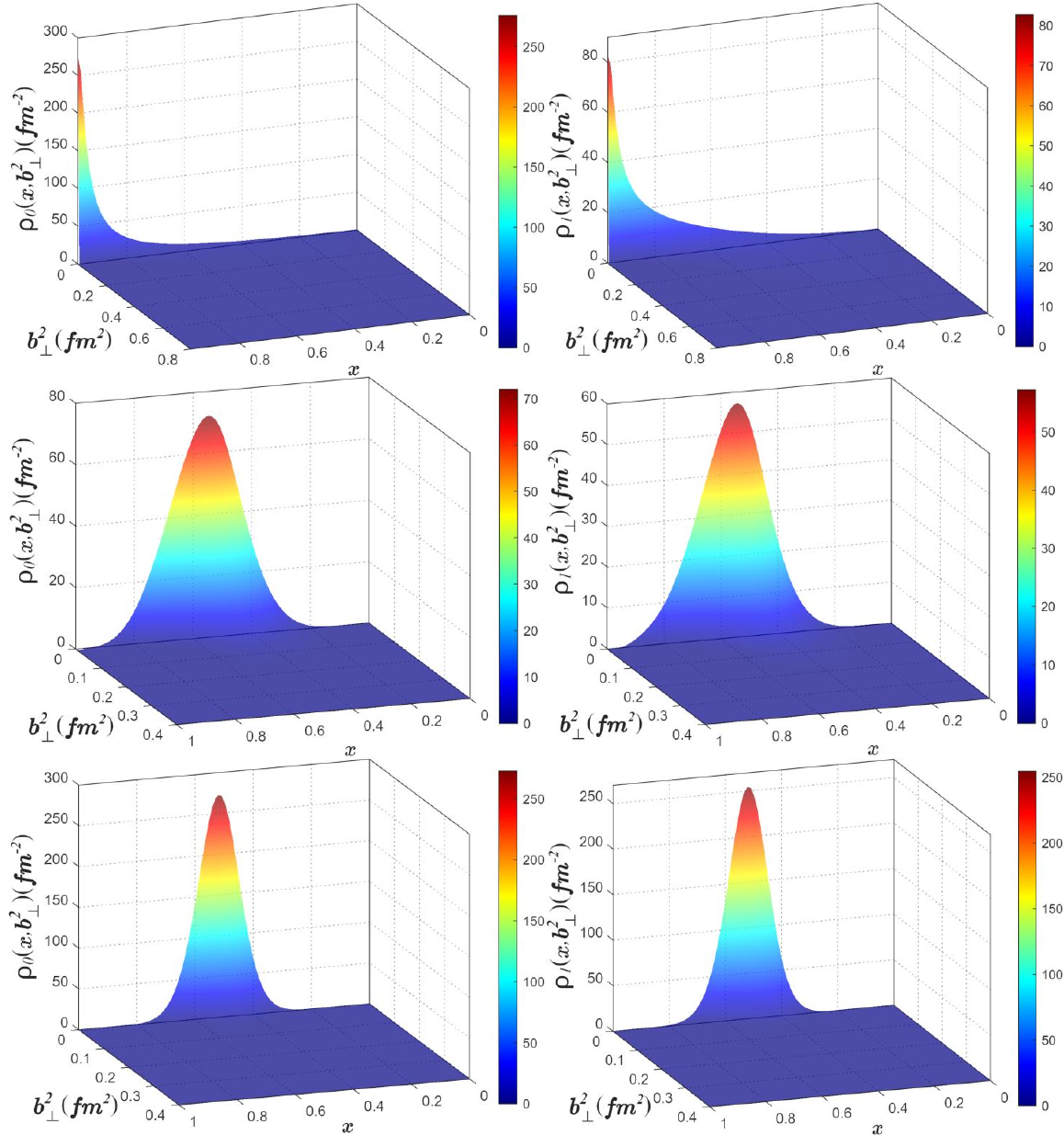} \\
\caption{\looseness=-1 
The IPD GPD $\rho_0(x,\vect{b}_\perp^2)$ (left column) and $\rho_1(x,\vect{b}_\perp^2)$ (right column) defined in Eq.~(\ref{eq:rho}) for $\rho$ (top row), $J/\psi$ (middle row) and $\Upsilon$ (bottom row) respectively. 
}
\label{fig:IPDGPD}
\end{figure}

\begin{figure}[htbp]
\centering\includegraphics[width=1.0\columnwidth]{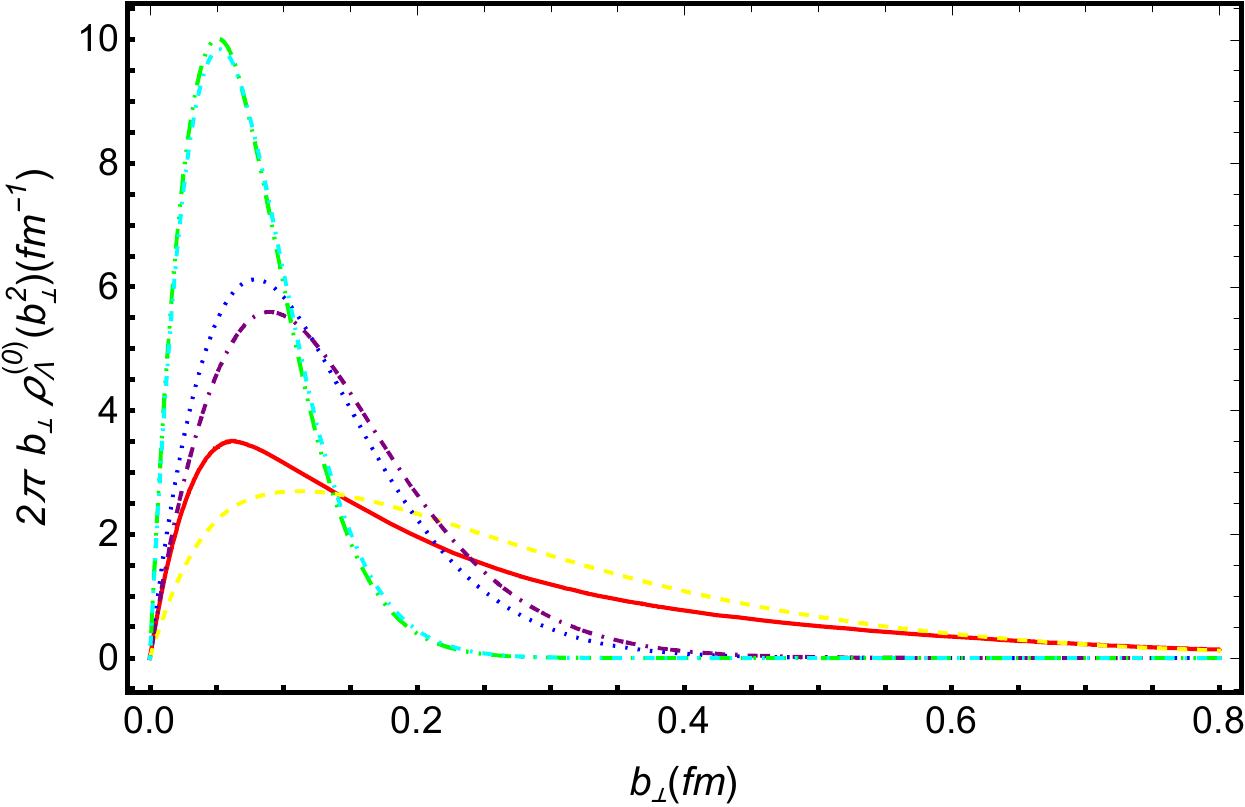} \\
\caption{\looseness=-1 
The sptial distribution of valence quarks of vector mesons.  The $\rho^{(0)}_0(\vect{b}_\perp^2)$ of $\rho$, $J/\psi$ and $\Upsilon$ are displayed in red solid, blue dotted and green dot-dot-dashed curves, and the $\rho^{(0)}_1(\vect{b}_\perp^2)$'s are displayed in yellow dashed, purple dot-dash-dashed and cyan  dot-dashed curves.
}
\label{fig:IPDGPD1D}
\end{figure}

Analogously, the Fourier transform of $V_{0,0}$ and $V_{1,1}$  can be interpreted as the unpolarized quark distribution inside helicity-0 and -1 vector mesons respectively, i.e.,
\begin{align} \label{eq:rho}
\rho_\Lambda(x,\vect{b}_\perp^2)=\int \frac{d^2 \vect{\Delta}_\perp}{(2\pi)^2} \textrm{e}^{-i \vect{b}_\perp \cdot \vect{\Delta}_\perp} V_{\Lambda, \Lambda}(x,0,-\vect{\Delta}_\perp^2). 
\end{align}
One also finds $\rho_{-1}(x,\vect{b}_\perp^2)=\rho_{1}(x,\vect{b}_\perp^2)$ according to Eq.~(\ref{eq:PI}).  In Fig.~\ref{fig:IPDGPD}, we show the $\rho_\Lambda(x,\vect{b}_\perp^2)$ for $\rho$, $J/\psi$ and $\Upsilon$ \footnote{The zero mode contribution is not taken into account in calculating the $\rho_{\Lambda}(x, \vect{b}_\perp^2)$, as it is essentially in the ERBL region and can not yield the probability density interpretation.}. Comparing the rows, one can see in heavier mesons, the quark distribution are more localized around $x=0.5$ and small $b_T$, indicating the heavy quark tend to carry half of the meson's longitudinal momentum and are spatially more centered. On the other hand, the $\rho_0(x,\vect{b}_\perp^2)$ and $\rho_1(x,\vect{b}_\perp^2)$ are a bit different by comparing the columns. To make it more transparent, we integrate over  $x$ and look into the spatial distribution 
\begin{align}
\rho^{(0)}_\Lambda(\vect{b}_\perp^2)\equiv \int_0^1 dx\rho_\Lambda(x,\vect{b}_\perp^2),
\label{eq:rho0}
\end{align}
which is displayed in Fig.~\ref{fig:IPDGPD1D}. We notice that the unpolarized quark are generally more broadly distributed in $\vect{b}_T$ in helicity-1 meson than in helicity-0 case. An underlying reason is that  the helicity-1 meson host components that have higher orbital angular momentum in the $z$-direction. For instance, at $q\bar{q}$ Fock-state truncation, the helicity-1 meson LF-LFWFs contain up to d-wave  components while in helicity-0 meson there are only s- and p-wave components \cite{Shi:2021taf,Shi:2022erw}. For the same reason, the difference between $\rho_0(x,\vect{b}_\perp^2)$ and $\rho_1(x,\vect{b}_\perp^2)$ significantly reduces in $J/\psi$ and $\Upsilon$, as p- and d-wave components are much more suppressed in heavy mesons. 

Finally we remind that the corresponding collinear unpolarized parton distribution functions of vector mesons have been reported in \cite{Shi:2022erw}, along with their transverse momentum dependent distributions. Therein we have determined the renormalization scale of our PDFs to be $\mu_0 \approx 670$ MeV, 2.6 GeV and 8.6 GeV for $\rho$, $J/\psi$ and $\Upsilon$ respectively. They should be considered to be the scale of our calculated GPDs herein as well.

\subsection{Electromagnetic form factors of vector mesons.}
\begin{figure}[htbp]
\centering\includegraphics[width=1.0\columnwidth]{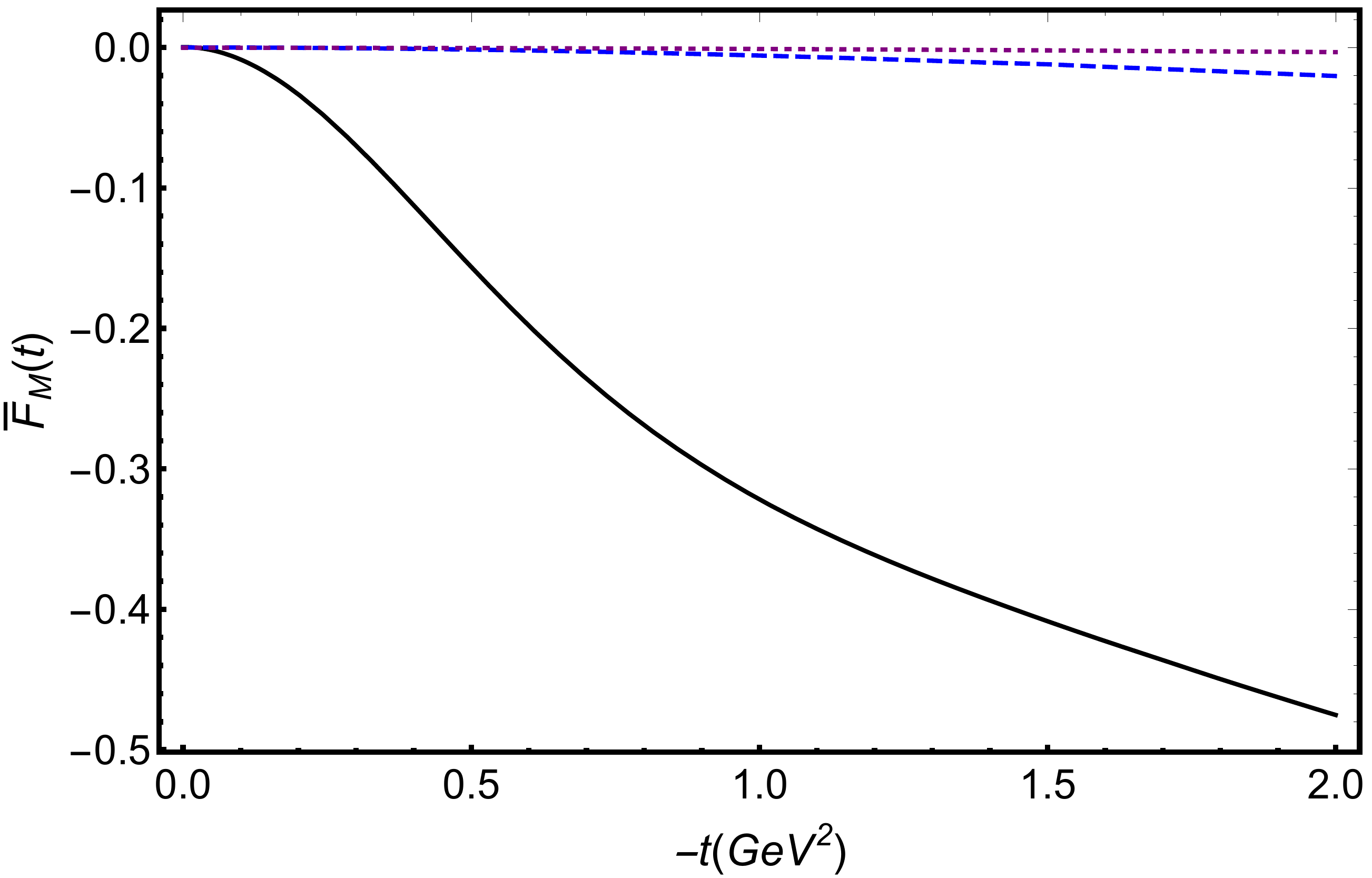} \\
\caption{\looseness=-1 
The zero mode contribution $\bar{F}_\rho(t)$ (black solid),  $\bar{F}_{J/\psi}(t)$ (blue dashed) and $\bar{F}_\Upsilon(t)$ (purple dotted) determined with angular momentum  condition Eq.~(\ref{eq:amc}). }
\label{fig:amc}
\end{figure}

Historically, the EMFFs are early and important tools to study the internal structure of hadrons. They enter the decomposition of the correlation function of current operator as
\begin{align}
G_{\Lambda',\Lambda}&=\frac{1}{2 P^+}\langle p',\Lambda'|\bar{\psi}(0)\gamma^+ \psi(0)|p,\Lambda \rangle \nonumber\\
&=-(\epsilon'^*\cdot \epsilon) F_1(t)+\frac{[\epsilon^+(\epsilon'^*\cdot P)+\epsilon'^{*+}(\epsilon \cdot P)]}{P^+}F_2(t) \nonumber \\
&\ \ -2\frac{(\epsilon\cdot P)(\epsilon'^* \cdot P)}{m_\rho^2}  F_3(t) \label{eq:EMFFdef}
\end{align} 
Comparing Eqs.~(\ref{eq:Vdef}, \ref{eq:Vdeco}) and Eq.~(\ref{eq:EMFFdef}), one finds the EMFFs are the first Mellin moments of the GPDs, i.e.,
\begin{equation}\label{eq:Fs}
F_i^q(t)=
\begin{cases}
\int_{-1}^1 dx H_i^q(x,\xi,t) \ ,\ \  i=1,2,3\\
0\ . \hspace{24mm} i=4,5
\end{cases}
\end{equation}
At this stage, one can resort to Eqs.~(\ref{eq:H1}-\ref{eq:H3}) and obtain the EMFFs. However, the $\bar{F}_M(t)$ in Eq.~(\ref{eq:Vfinal}) is not determined yet. Here, we determine $\bar{F}_M(t)$ using the angular momentum condition 
\begin{align}
(1+2\tau) G_{1,1}+G_{1,-1}-\sqrt{8\tau} G_{1,0}-G_{0,0}=0.\label{eq:amc}
\end{align}
This condition had been noticed in the study of vector meson EMFFs and physically it originates in angular momentum conservation \cite{Grach:1983hd,Choi:2004ww}.  It comes about as there are only three independent $F_i$'s in Eq.~(\ref{eq:EMFFdef}), hence the four $G_{\Lambda,\Lambda'}$'s (with different $\Lambda$ and $\Lambda'$) must be linearly dependent. It is equivalent to $F_5^q(t)=\int_{-1}^1 dx H_5(x,0,t)=0$ of Eqs.~(\ref{eq:Fs}), given Eq.~(\ref{eq:H5}) and $G_{\Lambda,\Lambda'}=\int_0^1 dx V_{\Lambda,\Lambda'}(x,0,t)$. Here we remark that the modification term $\tilde{F}_M(t)$ in Eq.~(\ref{eq:Vfinal}), which is associated with the second kind of zero mode, only brings an overall multiplicative factor $1+\tilde{F}_M(t)$ to the left hand side of Eq.~(\ref{eq:amc}), hence can not fix the angular momentum condition. While the $\bar{F}_M(t)$ in Eq.~(\ref{eq:Vfinal}) adds an inhomogeneous term to the left hand side of Eq.~(\ref{eq:amc}) so it is indispensable to fix the issue. In Fig.~\ref{fig:amc} we plot the $\bar{F}_\rho(t)$, $\bar{F}_{J/\psi}(t)$ and $\bar{F}_\Upsilon(t)$ determined from Eq.~(\ref{eq:amc}). We notice they are significantly suppressed in the heavy sector. With the angular momentum condition respected, different prescriptions for EMFFs, such as the Grach and Kondratyuk (GK) \cite{Grach:1983hd} and Brodsky and Hiller (BH) \cite{Brodsky:1992px} prescriptions, are now equivalent and free of theoretical ambiguity.

\begin{figure}[htbp]
\centering\includegraphics[width=1.0\columnwidth]{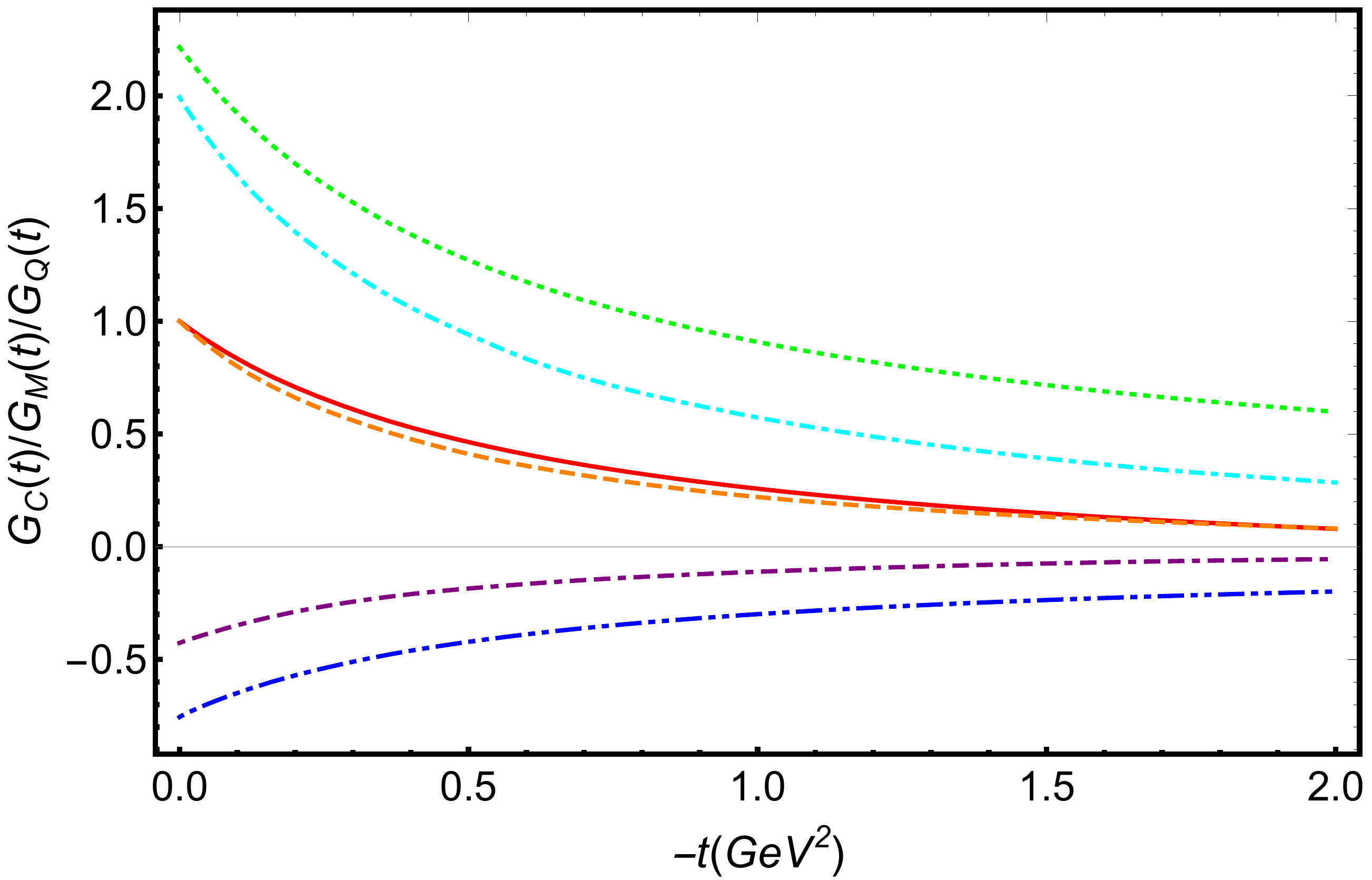} \\
\caption{\looseness=-1 
The EMFFs of $\rho$. Our calculated $G_C(t)$, $G_M(t)$ and $G_Q(t)$ are displayed in red solid, green dotted and blue dot-dot-dashed curves. The fully covariant Bethe-Salpter approach with Maris-Tandy model gives the orange dashed, cyan dot-dashed and purpule dot-dash-dashed curves respectively  \cite{Bhagwat:2006pu}. }
\label{fig:rhoEMFF}
\end{figure}

\begin{figure}[htbp]
\centering\includegraphics[width=1.0\columnwidth]{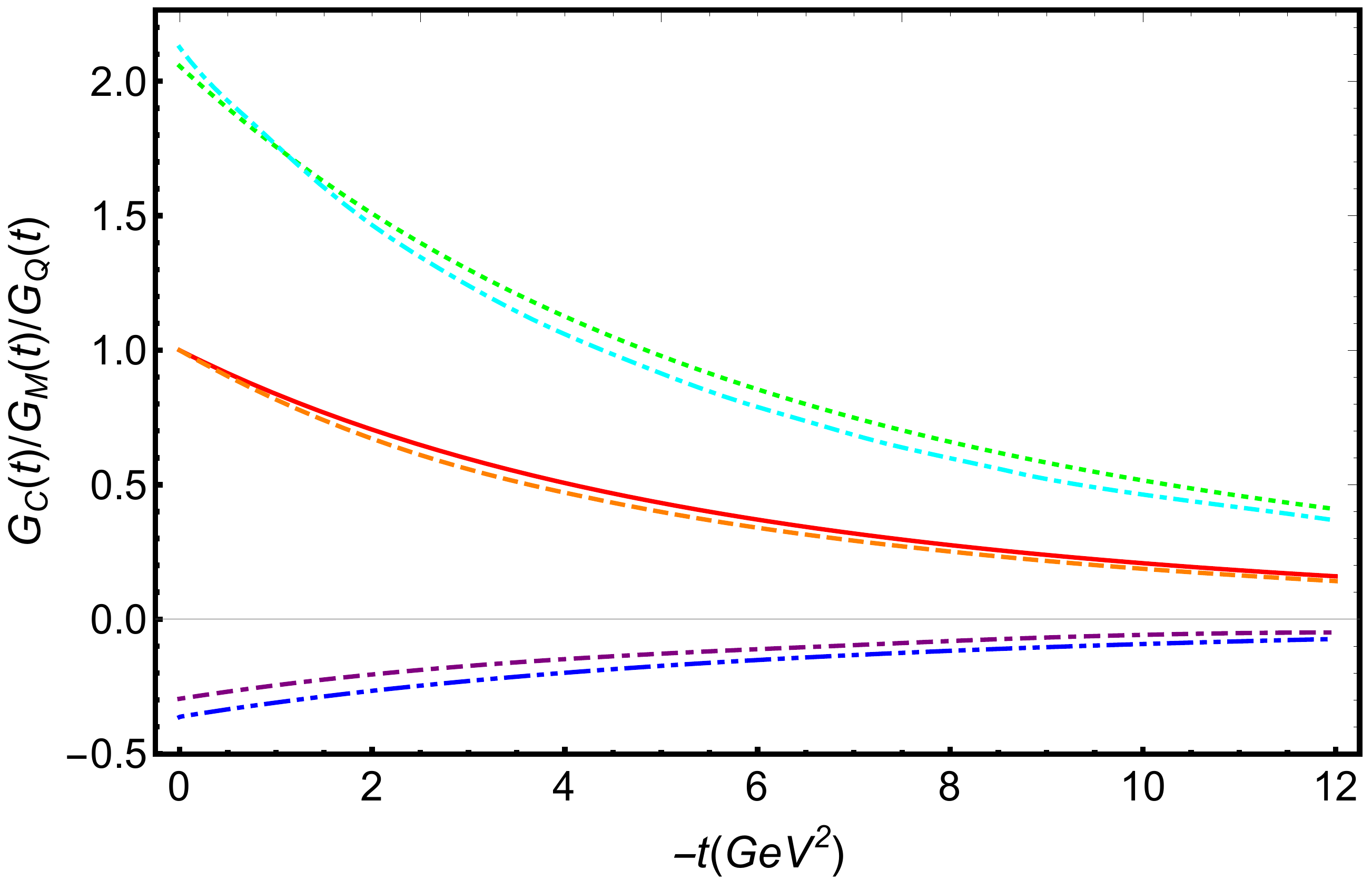} \\
\caption{\looseness=-1 
The EMFFs of $J/\psi$. Our calculated $G_C(t)$, $G_M(t)$ and $G_Q(t)$ are displayed in red solid, green dotted and blue dot-dot-dashed curves. The fully covariant Bethe-Salpter approach with Maris-Tandy model gives the orange dashed, cyan dot-dashed and purpule dot-dash-dashed curves respectively \cite{Maris:2006ea}. }
\label{fig:JpsiEMFF}
\end{figure}

\begin{figure}[htbp]
\centering\includegraphics[width=1.0\columnwidth]{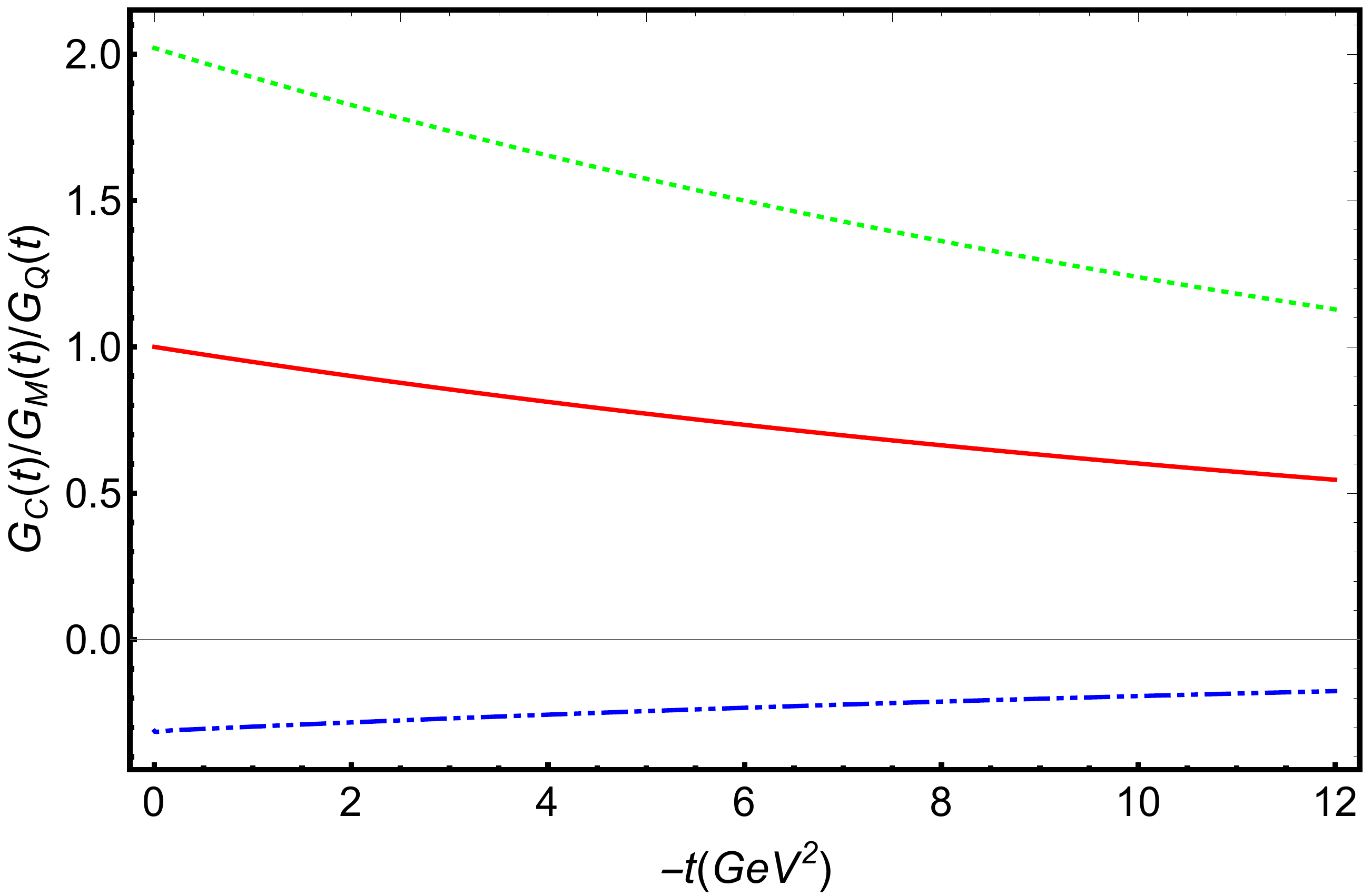} \\
\caption{\looseness=-1 
The $G_C(t)$, $G_M(t)$ and $G_Q(t)$ for $\Upsilon$ displayed in red solid, green dotted and blue dot-dot-dashed curves. }
\label{fig:UpsilonEMFF}
\end{figure}

The charge, magnetic and quadrupole form factors of vector mesons are related to the form factors $F_i$ by
\begin{align}
G_C(t)&=\left(1+\frac{2}{3} \tau\right)F_1(t)+\frac{2}{3} \tau F_2(t)+\frac{2}{3}\tau\left(1+\tau\right)F_3(t) \\
G_M(t)&=-F_2(t) \\
G_Q(t)&=F_1(t)+F_2(t)+(1+\tau)F_3(t)
\end{align} 
We plot them for  $\rho$, $J/\psi$ and $\Upsilon$ in Figs.~\ref{fig:rhoEMFF}, \ref{fig:JpsiEMFF} and \ref{fig:UpsilonEMFF} respectively. Since the $J/\psi$ and $\Upsilon$ are electric neutral, their results only take the valence quark contribution to EMFFs into account, as was done in \cite{Maris:2006ea,Adhikari:2018umb}.  We also supplement with the fully covariant DS-BSEs calculation on $\rho$ and $J/\psi$ EMFFs in the plots for comparison \cite{Maris:2006ea,Bhagwat:2006pu}. From Fig.~\ref{fig:rhoEMFF} we can see that for the light meson $\rho$, the two approaches yield results that are somewhat different. This is mainly due to the leading Fock-state truncation we impose. While for heavier meson $J/\psi$, the agreement is much better and extends into the high $Q^2$ region, suggesting the leading Fock-state truncation to be more valid in heavy mesons.

From the plots we extract the charge radius $\langle r^2 \rangle$, the magnetic
moment $\mu$, and the quadrupole moment $Q$ of the vector mesons, which are defined as 
\begin{align}
\langle r^2\rangle_c&=\left. 6\frac{\partial}{\partial t} G_c(t)\right |_{t\rightarrow 0}, \\
\mu&=G_M(0)\times \frac{1}{2 m_V},\\
Q&=G_Q(0)\times \frac{1}{m^2_V}.
\end{align}
Our calculated values are listed in the first three columns, with subscript LFBS referring to a combined effort of light front approach and the Bethe-Salpeter approach. Our $\rho$ charge radius is smaller than the full Bethe-Salpeter equation and lattice calculation, but larger than other light front approach results \cite{DeMelo:2018bim, Qian:2020utg}. If we remove the modification term associated with $\tilde{F}_M$ in Eq.~(\ref{eq:Vfinal}), our result would be close to theirs. Similar to \cite{DeMelo:2018bim, Qian:2020utg}, our $\rho$ magnetic  and  quadrupole moments are generally larger in magnitude than the full BSE calculation, but somehow closer to the lattice prediction \cite{Owen:2015gva}. Meanwhile, there is better agreement between LFBS, full BSE and lattice calculations on $J/\psi$. We also notice a monotonic decrease in the magnitude of $\langle r^2\rangle$, $\mu$ and $Q$ from $\rho$ to $J/\psi$ and eventually $\Upsilon$.

\begin{table*}[htbp]
	\label{tab:rmuQ}
	\begin{center}
\begin{tabular}{cccccccccccc}
	\toprule
	\toprule
 & $\rho_{\rm LFBS}$  \ \ \ & $J/\psi_{\rm LFBS}$  \ \ \  & $\Upsilon_{\rm LFBS}$  \ \ \  & $\rho_{\rm BS}$ \cite{Bhagwat:2006pu} \ \ \  & $J/\psi_{\rm BS}$ \cite{Bhagwat:2006pu} \ \ \  & $\rho_{\rm LF1} $ \cite{DeMelo:2018bim} \ \ \  &  $\rho_{\rm LF2}$ \cite{Choi:2004ww}  &  $\rho_{\rm LF3}$ \cite{Qian:2020utg}\ \ \  & $\rho_{\rm Lat}$   \cite{Owen:2015gva}\ \ \  & $J/\psi_{\rm Lat}$  \cite{Dudek:2006ej} \ \ \  \\
    \midrule
$\sqrt{\langle \vect{r}^2  \rangle_c}$ (fm)& 0.66 & 0.21 & 0.11 & 0.74 & 0.23 & 0.52 & -& 0.48  &0.819(43) & 0.257(4) \\
$\mu \times 2 m_V$ &2.21 & 2.06 & 2.02 & 2.01 & 2.13 &2.10 & 1.92 & 2.15 &2.209(82) & 2.10(3) \\
$Q \times m^2_V$ & -0.76& -0.36 & -0.31 & -0.41 &-0.28(1) & -0.898 &-0.43&-0.886 & -0.733(99) &-0.23(2) \\
   \bottomrule
   \bottomrule
   \end{tabular}
\end{center}
\caption{The charge radius and magnetic multipole moments of vector mesons. Our results are in the first three columns. The lattice simulation for $\rho$ meson is performed at $\rho$ mass of 793 MeV \cite{Owen:2015gva}.}
\end{table*}

\subsection{Gravitational form factors of vector mesons.}

The GFFs are defined as the form factors in the decomposition of the energy-momentum tensor of QCD , i.e., $  \langle p', \Lambda' | T_{\mu\nu}(0) | p, \Lambda \rangle$. At present, there are different definitions of the QCD's EMT operator $T_{\mu\nu}$. The Belinfante-Rosenfeld EMT, for instance, is symmetric in $\mu$ and $\nu$ \cite{belinfante1939spin,rosenfeld1940energy}, and the canonical EMT or the gauge-invariant kinetic
EMT \cite{Leader:2013jra, Cosyn:2019aio} are not. Here we take the Belinfante-Rosenfeld EMT, which reads 
\begin{align}
T_{\mu \nu}&=T^q_{\mu \nu}+T^g_{\mu \nu}, \\
T^q_{\mu \nu}&=\frac{1}{4}\bar{\psi}_q\left(\gamma_\mu i \overset{\! \! \leftrightarrow}{D_\nu}+\gamma_\nu i \overset{\!\! \leftrightarrow}{D_\mu}\right)\psi_q-g_{\mu \nu}\bar{\psi}_q\left(\frac{i}{2}\overset{\! \! \leftrightarrow}{\slashed{D}}-m\right)\psi_q, \\
T^g_{\mu\nu}&=F^{a,}_{\ \ \mu \rho} F^{a,\rho}  \ _{\! \! \! \nu}+\frac{1}{4}g_{\mu\nu} F^{a,\rho\lambda} F^{a,}_{\ \ \rho \lambda}.
\end{align}
with $\overset{\!\! \leftrightarrow}{D_\mu}=(\overset{\!\! \rightarrow}{\partial_\mu}-\overset{\!\! \leftarrow}{\partial_\mu})-2 i g t^a A_\mu^a$ and $F^{a}_{\ \mu \nu}=\partial_\mu A_\nu^a-\partial_\nu A^a_\mu+g f^{abc}A^b_\mu A^c_\nu$, whose decomposition has thus only symmetric terms \cite{Holstein:2006ge,Abidin:2008ku,Taneja:2011sy,Cosyn:2019aio,Polyakov:2019lbq}, i.e.,
\begin{widetext}
\begin{align}
  \langle p', \Lambda' | T_{\mu\nu}(0) | p, \Lambda \rangle & = -2P_\mu P_\nu\left[(\epsilon'^{*} \epsilon)\mathcal{G}_1(t) -\frac{(\Delta\epsilon'^{*} )(\Delta\epsilon )}{2m_\rho^2} \mathcal{G}_2(t)\right]- \frac{1}{2}(\Delta_\mu \Delta_\nu - \Delta^2 g_{\mu\nu})\left[(\epsilon'^{*} \epsilon)\mathcal{G}_3(t) -\frac{(\Delta\epsilon'^{*} )(\Delta\epsilon )}{2m_\rho^2}
  \mathcal{G}_4(t)\right]\nonumber\\
&  +P_{\{\mu}\left( \epsilon'^{*}_{\nu\}} (\Delta \epsilon)- \epsilon_{\nu\}} (\Delta \epsilon'^{*}) \right)\mathcal{G}_5(t)+ \frac{1}{2} \left[\Delta_{\{\mu}\left( \epsilon'^{*}_{\nu\}} (\Delta\epsilon)+ \epsilon_{\nu\}} (\Delta\epsilon'^{*}) \right)- \epsilon_{\{\mu}'^{*}\epsilon_{\nu\}} \Delta^2- g_{\mu\nu}(\Delta\epsilon'^{*})(\Delta\epsilon)\right]\mathcal{G}_6(t), \label{eq:fullEMT}
\end{align}
where $A_{[ \mu} B_{\nu ]} = (A_\mu B_\nu-A_\nu B_\mu)/2$
and $A_{\{\mu} B_{\nu\}} = (A_\mu B_\nu+A_\nu B_\mu)/2$. The notation for GFFs here follows that in \cite{Cosyn:2019aio}, and its equivalence to other notations is summarized in \cite{Polyakov:2019lbq}. More generally, the quark or gluon EMT doesn't have to be conserved separately, so it has three additional symmetric tensor structures \cite{Cosyn:2019aio}
\begin{align}
  \langle p', \Lambda' | T^a_{\mu\nu}(0) | p, \Lambda \rangle & = -2P_\mu P_\nu\left[(\epsilon'^{*} \epsilon)\mathcal{G}^a_1(t) \frac{(\Delta\epsilon'^{*} )(\Delta\epsilon )}{2m_\rho^2} -\mathcal{G}^a_2(t)\right]- \frac{1}{2}(\Delta_\mu \Delta_\nu - \Delta^2 g_{\mu\nu})\left[(\epsilon'^{*} \epsilon)\mathcal{G}^a_3(t) -\frac{(\Delta\epsilon'^{*} )(\Delta\epsilon )}{2m_\rho^2}
  \mathcal{G}^a_4(t)\right]\nonumber\\
&  +P_{\{\mu}\left( \epsilon'^{*}_{\nu\}} (\Delta \epsilon)- \epsilon_{\nu\}} (\Delta \epsilon'^{*}) \right)\mathcal{G}^a_5(t)+ \frac{1}{2} \left[\Delta_{\{\mu}\left( \epsilon'^{*}_{\nu\}} (\Delta\epsilon)+ \epsilon_{\nu\}} (\Delta\epsilon'^{*}) \right)- \epsilon_{\{\mu}'^{*}\epsilon_{\nu\}} \Delta^2- g_{\mu\nu}(\Delta\epsilon'^{*})(\Delta\epsilon)\right]\mathcal{G}^a_6(t)\nonumber \\ 
&+\epsilon_{\{\mu}'^{*}\epsilon_{\nu\}} m_\rho^2 \mathcal{G}^a_7(t)+ g_{\mu\nu} m_\rho^2 (\epsilon'^*\epsilon) \mathcal{G}^a_8(t)+ \frac{1}{2}g_{\mu\nu}(\Delta\epsilon'^*)( \Delta\epsilon) \mathcal{G}^a_9(t),\label{eq:aEMT}
\end{align}
\end{widetext}
The superscript $a$ could be either quark or gluon. Given that $T_{\mu\nu}(0)=\sum_a T^a_{\mu\nu}(0)$, one has $\mathcal{G}_i(t)=\sum_a \mathcal{G}^a_i(t)$. Due to the leading Fock-state truncation we employ, the gluon EMT FFs vanish. Consequently the quark contribution $\mathcal{G}^q_i, i\in\{7,8,9\}$ vanish as $\sum_a \mathcal{G}^a_i(t)=0, i\in\{7,8,9\}$. 

\begin{figure}[htbp]
\centering\includegraphics[width=1.0\columnwidth]{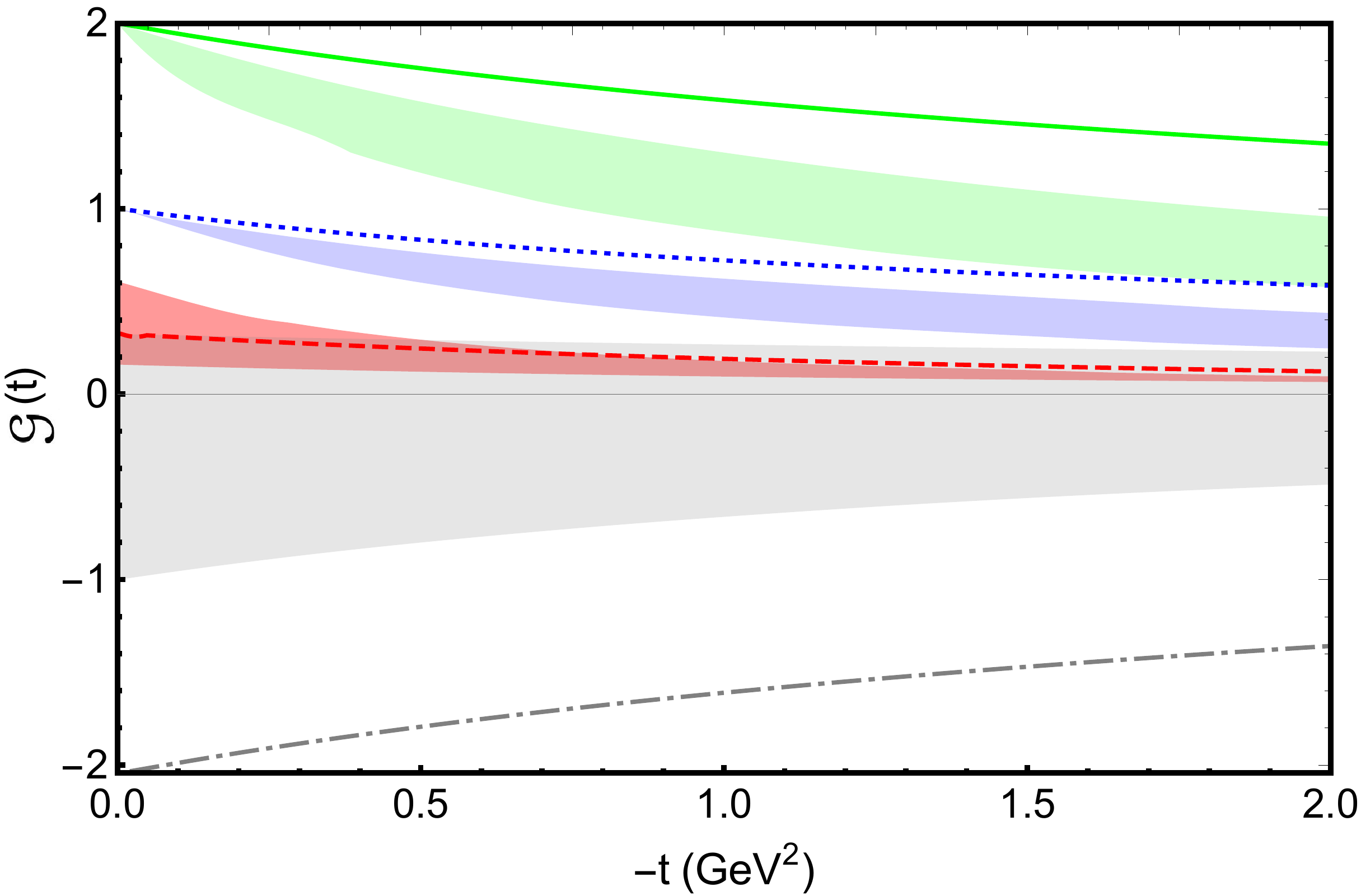} \\
\caption{\looseness=-1 
Our calculated gravitational form factors $\mathcal{G}_1$ (blue dotted), $\mathcal{G}_2$ (red dashed), $\mathcal{G}_5$ (green solid), $\mathcal{G}_6$ (gray dot-dashed) of $\rho$ displayed in curves. The colored bands are enveloped by NJL model \cite{Freese:2019bhb} and LFCQM \cite{Sun:2020wfo}. At $t\approx 0$ GeV$^2$, from top to bottom, these bands correspond to $\mathcal{G}_5$, $\mathcal{G}_1$, $\mathcal{G}_2$ and $\mathcal{G}_6$ respectively. Among them, the NJL model yields the upper boundaries of  $\mathcal{G}_5$ and $\mathcal{G}_1$, and the lower  boundaries of $\mathcal{G}_2$ and $\mathcal{G}_6$.}
\label{fig:RhoGFF}
\end{figure}

\begin{figure}[htbp]
\centering\includegraphics[width=1.0\columnwidth]{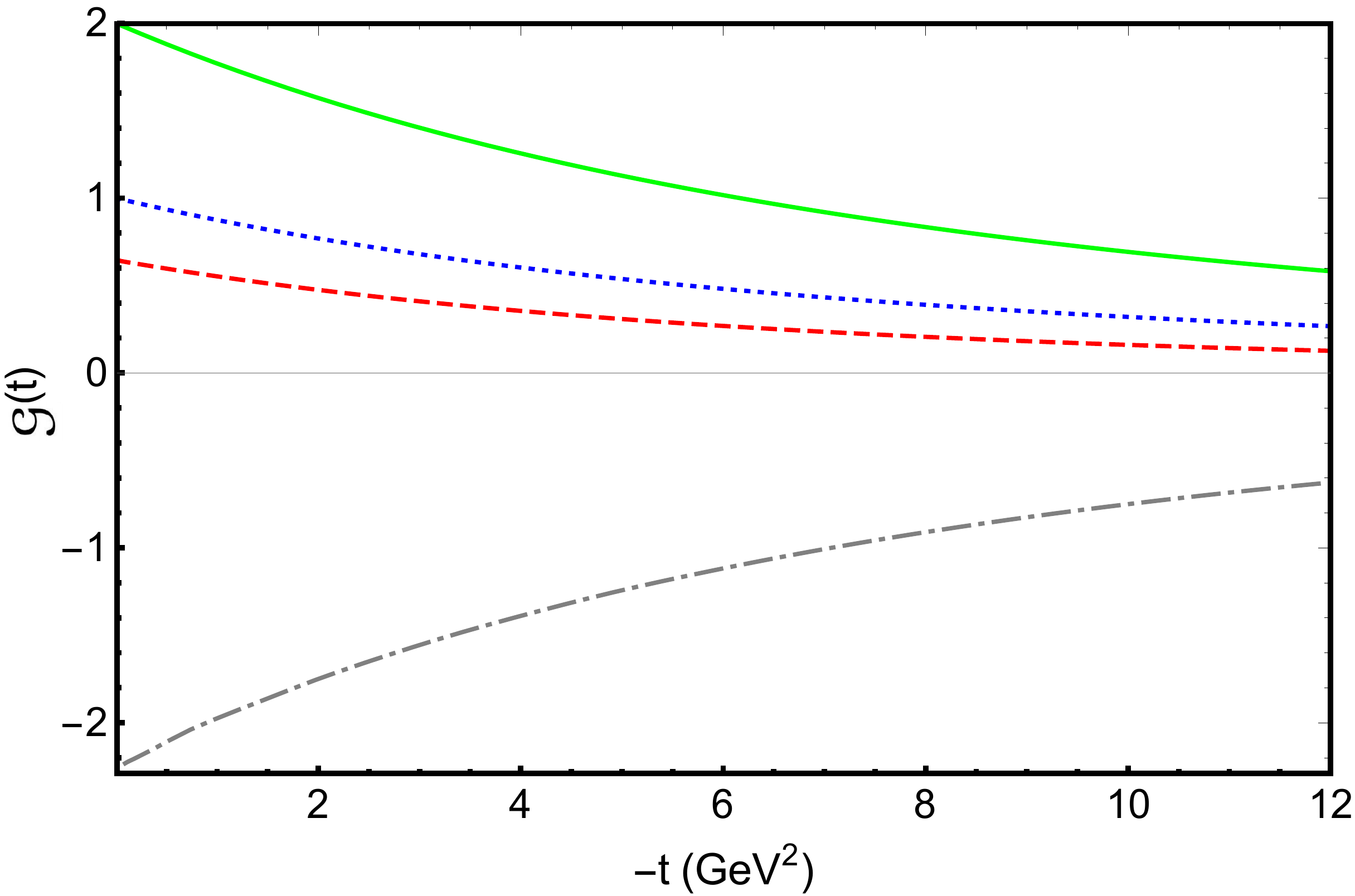} \\
\caption{\looseness=-1 
The gravitational form factors $\mathcal{G}_1$ (blue dotted), $\mathcal{G}_2$ (red dashed), $\mathcal{G}_5$ (green solid), $\mathcal{G}_6$ (gray dot-dashed) of $J/\psi$. }
\label{fig:JpsiGFF}
\end{figure}

\begin{figure}[htbp]
\centering\includegraphics[width=1.0\columnwidth]{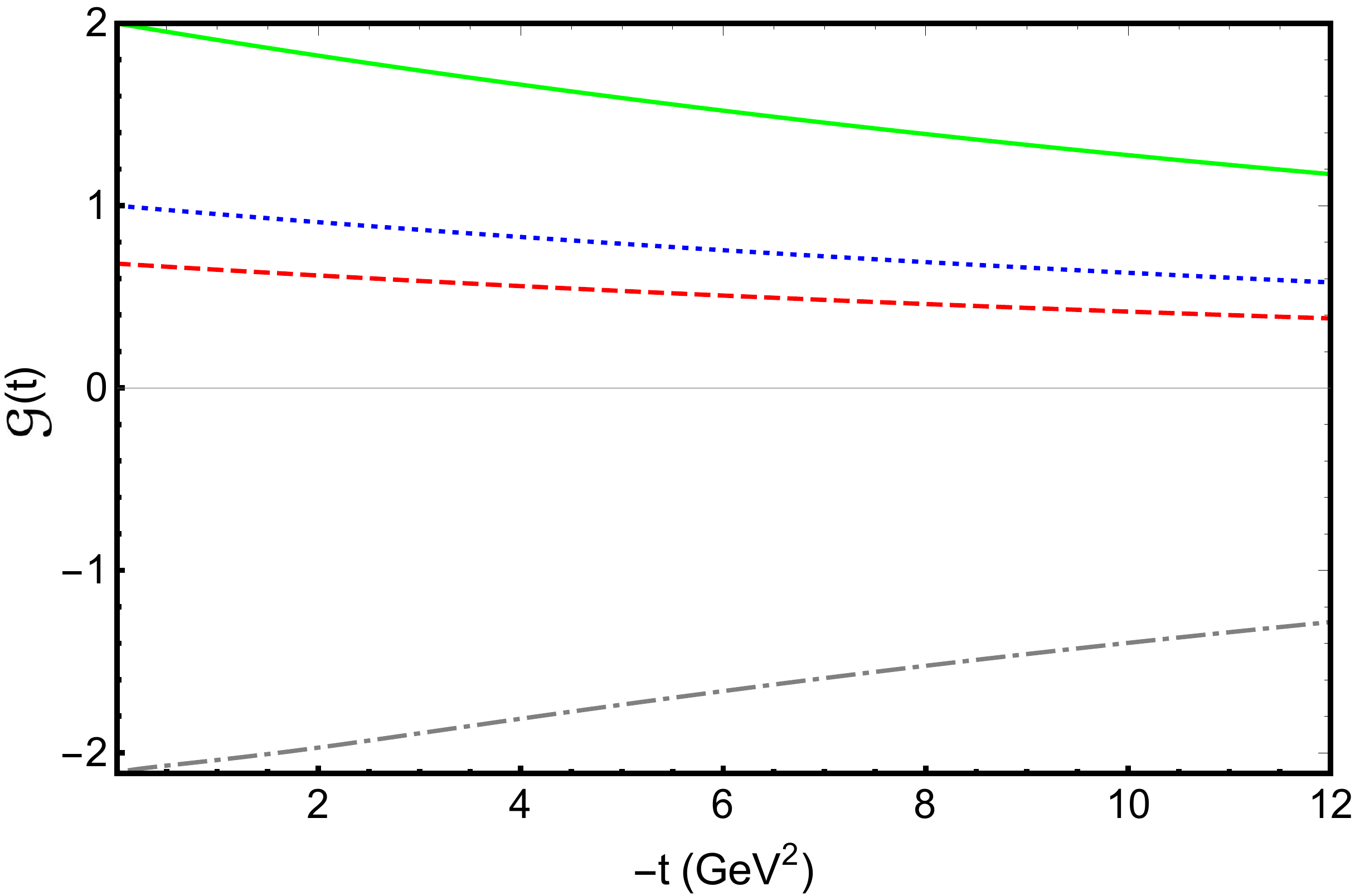} \\
\caption{\looseness=-1 
The gravitational form factors $\mathcal{G}_1$ (blue dotted), $\mathcal{G}_2$ (red dashed), $\mathcal{G}_5$ (green solid), $\mathcal{G}_6$ (gray dot-dashed) of $\Upsilon$. }
\label{fig:UpsilonGFF}
\end{figure}

In experiment, the GFFs are not directly measurable, but rather connected with the GPDs. Comparing their definitions, the quark GFFs can be connected with the second  Mellin moments of the  unpolarized  quark GPDs \cite{Abidin:2008ku,Taneja:2011sy,Cosyn:2019aio,Polyakov:2019lbq}
\begin{align}
 \int_{-1}^1\mathrm{d}x\, x \left[ H^{a}_1(x,\xi,t) - \frac{1}{3}H^{q}_5(x,\xi,t)\right] 
  &= \mathcal{G}^{q}_1(t) + \xi^2 \mathcal{G}^{q}_3(t), \label{eq:GPD2GFF1}\\
 \int_{-1}^1\mathrm{d}x\, x H^{q}_2(x,\xi,t)  &= \mathcal{G}^{q}_5(t), \\
 \int_{-1}^1\mathrm{d}x\, x H^{q}_3(x,\xi,t)  &= \mathcal{G}^{q}_2(t)
  + \xi^2 \mathcal{G}^{q}_4(t), \\
 \int_{-1}^1\mathrm{d}x\, x H^{q}_4(x,\xi,t)  &= \xi \mathcal{G}^{q}_6(t),\\
  \int_{-1}^1\mathrm{d}x\, x H^{q}_5(x,\xi,t)  &= -\frac{t}{4M^2}\mathcal{G}^{q}_6(t)
  + \frac{1}{2}\mathcal{G}^{q}_7(t).
  \label{eq:GPD2GFF5}
\end{align}
Since our calculated GPDs are limited to zero skewness, four GFFs, e.g., the $\mathcal{G}^q_1$, $\mathcal{G}^q_2$, $\mathcal{G}^q_5$ and $\mathcal{G}^q_6$ can be extracted. In Figs.~\ref{fig:RhoGFF}-\ref{fig:UpsilonGFF}, we display them for $\rho$, $J/\psi$ and $\Upsilon$ respectively. In Fig.~\ref{fig:RhoGFF}, we display our calculated $\rho$ GFFs as curves, while the colored bands are enveloped by results from NJL model \cite{Freese:2019bhb} and LFCQM model \cite{Sun:2020wfo}. We only show results up to 2 GeV$^2$, as the high-$t$ region should be dominated by LF-LFWFs \emph{before} the rescaling procedure in Eq.~(\ref{eq:N2}) \cite{Tong:2022zax}. With the momentum sum rule $\mathcal{G}_1(0)=1$ and angular momentum sum rule \cite{Taneja:2011sy} $\mathcal{G}_5(0)=2$ automatically hold, our results are generally closer to the NJL model prediction, including the sign of the $g_6(t)$.  However, our $g_6(t)$ is significantly larger in magnitude, i.e., it is twice that of NJL model at the origin. For the heavy mesons, we predict their GFFs up to 12 GeV$^2$, as displayed in Figs.~\ref{fig:JpsiGFF} and \ref{fig:UpsilonGFF}. From these GFFs, one can extract the light front mass radii of vector mesons through \cite{Freese:2019bhb} 
\begin{align}
\langle \vect{r}_\perp^2 \rangle_{\textrm{LC}}&\equiv \lim_{\vect{\Delta}_\perp\rightarrow 0}-\frac{1}{P^+}\nabla^2_{\vect{\Delta}_\perp}\left[ \left . \frac{1}{2P^+}\langle p',\Lambda|T^{++}(0)| p,\Lambda \rangle \right |_{\Delta^+=0} \right]\\
&=4 \frac{d \mathcal{G}_1(t)}{dt}+\frac{1}{m_V}\left[ \frac{2}{3}\mathcal{G}_1(0)-\frac{2}{3}\mathcal{G}_2(0) \right. \nonumber\\
&\left. \hspace{10mm} -\frac{1}{3}\mathcal{G}_5(0)-\frac{1}{3}\mathcal{G}_6(0)\right].
\end{align}
We find $\sqrt{\langle \vect{r}_\perp^2 \rangle^\rho_{\textrm{LC}}}=0.30$ fm, which is comparable to the NJL model result 0.32 fm \cite{Freese:2019bhb} and LFQCM model result 0.41 fm \cite{Sun:2020wfo}. We also find $\sqrt{\langle \vect{r}_\perp^2 \rangle^{J/\psi}_{\textrm{LC}}}=0.151$ fm and $\sqrt{\langle \vect{r}_\perp^2 \rangle^{\Upsilon}_{\textrm{LC}}}=0.087$ fm, showing the heavy mesons are  spatially more compact in energy distribution. We notice these values are almost identical to what we found for pseudoscalar mesons, i.e., $\sqrt{\langle \vect{r}_\perp^2 \rangle^{\eta_c}_{\textrm{LC}}}=0.150$ fm and $\sqrt{\langle \vect{r}_\perp^2 \rangle^{\eta_b}_{\textrm{LC}}}=0.089$ fm, with exactly same DS-BSEs interaction models \cite{Shi:2021nvg}. 

\section{Summary\label{sec:sum}}

The GPDs of light and heavy vector mesons, i.e. the $\rho$, $J/\psi$ and $\Upsilon$, at zero skewness are investigated with a combined effort from the light front and DSEs framework. Potential zero mode contributions are considered, and the  light front overlap representation is revised with an ansatz, e.g., Eq.~(\ref{eq:Vfinal}). Vector meson LF-LFWFs determined from DS-BSEs approach are then employed to study the GPDs.

As collinear parton distributions had been reported in \cite{Shi:2022erw}, in this work we focus on the 3-dimensional distribution IPD GPD. We show that the valence quark distributions are spatially broader in transversely polarized vector mesons ($|\Lambda|=1$) than in longitudinal mesons ($\Lambda=0$). We argue  this is because there are more Fock components with higher orbital angular momentum in transversely polarized mesons. This is  supported by our further finding that the difference between $\rho_0$ and $\rho_1$ are significantly reduced in heavy mesons, which are s-wave dominated systems.

We then investigate the EMFFs of the vector mesons. The zero mode contributions of Eq.~(\ref{eq:Vfinal}) play important roles in this case. For instance, the $\bar{F}_M(t)$ term restores the angular momentum condition Eq.~(\ref{eq:amc}) and removes the ambiguity in calculating the EMFFs. In this regard, such revision is necessary for theoretical consistency in modeling the GPDs. The other zero mode contribution, which arises from the dressing of quark-photon vertex, softens the EMFFs. Namely, it makes the EMFFs decrease faster and yields a larger charge radius.  Before the introduction of GPDs, such contribution was intuitively interpreted as the form factor of a parton-like quark inside a constituent quark \cite{Cardarelli:1994ix}. By comparing the obtained $\rho$ and $J/\psi$ EMFFs with fully covariant DS-BSEs calculation \cite{Maris:2006ea,Bhagwat:2006pu}, we notice the agreement gets much improved from $\rho$ to $J/\psi$. We therefore consider the deviations resides in the leading Fock-state truncation, which works much better for heavy systems.  

The gravitational form factors of the vector mesons are finally studied. As the second Mellin moments of GPDs, the GFFs receive no contribution from the zero mode, which is different from the EMFFs. In the leading Fock-state approximation, the GFFs come solely from the quarks, and certain GFFs can be extracted. Our $\rho$ GFFs are shown in Fig.~\ref{fig:RhoGFF} and compared with NJL model and LFCQM predictions. We also predict the GFFs for $J/\psi$ and $\Upsilon$, which had not been reported in the literatures before. Based on the experience from EMFFs, we believe they should be very close to a fully covariant calculation, which remains to be checked by DS-BSEs or other models in the future.  

\begin{acknowledgments}
This work is supported by the National Natural Science Foundation of China (under Grant No. 11905104).
\vspace{3em}
\end{acknowledgments}

\bibliography{ VGPD}

\end{document}